\documentclass[apj]{emulateapj}

\usepackage{graphicx}
\usepackage{epsfig}
\usepackage{natbib}
\newcommand{\fesc}{\ifmmode{f_{\rm esc}}\else{$f_{\rm esc}$}\fi}
\newcommand{\fescs}{\ifmmode{f_{\rm esc}^\star}\else{$f_{\rm esc}^\star$}\fi}
\newcommand{\kms}{\ifmmode{{\rm km~s^{-1}}}\else{km s$^{-1}$}\fi}
\newcommand{\cubecm}{\ifmmode{{\rm cm^{-3}}}\else{cm$^{-3}$}\fi}
\newcommand{\lsim}{\lower0.3em\hbox{$\,\buildrel <\over\sim\,$}}
\newcommand{\gsim}{\lower0.3em\hbox{$\,\buildrel >\over\sim\,$}}

\newcommand{\sfr}{M$_\odot$ yr$^{-1}$ Mpc$^{-3}$}

\newcommand{\enzo}{{\sl Enzo}}
\newcommand{\Ms}{\ifmmode{M_\odot}\else{$M_\odot$}\fi}

\newcommand{\hh}{H$_2$}

\newcommand{\tvir}{\ifmmode{T_{\rm{vir}}}\else{$T_{\rm{vir}}$}\fi}
\newcommand{\mvir}{\ifmmode{M_{\rm{vir}}}\else{$M_{\rm{vir}}$}\fi}
\newcommand{\rvir}{\ifmmode{r_{\rm{vir}}}\else{$r_{\rm{vir}}$}\fi}

\newcommand{\lya}{Ly$\alpha$}
\newcommand{\jj}{\ifmmode{J_{21}}\else{$J_{21}$}\fi}
\newcommand{\flw}{\ifmmode{F_{LW}}\else{$F_{LW}$}\fi}
\newcommand{\kph}{\ifmmode{k_{\rm ph}}\else{$k_{\rm ph}$}\fi}

\newcommand{\zsun}{{\rm\,Z_\odot}}

\newcommand\tento[1]{$10^{#1}$}

\newcommand{\ztwo}{\ifmmode{{\rm [Z_2/H]}}\else{[Z$_2$/H]}\fi}
\newcommand{\zthree}{\ifmmode{{\rm [Z_3/H]}}\else{[Z$_3$/H]}\fi}

\begin{document}

\shorttitle{THE BIRTH OF A GALAXY}
\shortauthors{WISE ET AL.}

\title{The Birth of a Galaxy: Primordial Metal Enrichment and Stellar
  Populations}

\author{John H. Wise\altaffilmark{1}, 
  Matthew J. Turk\altaffilmark{2},
  Michael L. Norman\altaffilmark{3},
  Tom Abel\altaffilmark{4,5,6}}

\altaffiltext{1}{Center for Relativistic Astrophysics, School of
  Physics, Georgia Institute of Technology, 837 State Street, Atlanta,
  GA 30332}
\altaffiltext{2}{Department of Astronomy, Columbia University, 538
  West 120th Street, New York, NY 10027}
\altaffiltext{3}{Center for Astrophysics and Space Sciences,
  University of California at San Diego, La Jolla, CA 92093}
\altaffiltext{4}{Kavli Institute for Particle Astrophysics and
  Cosmology, Stanford University, Menlo Park, CA 94025}
\altaffiltext{5}{Zentrum f\"ur Astronomie der Universit\"at, Institut
  f\"ur Theoretische Astrophysik, Albert-Ueberle-Str. 2, 69120
  Heidelberg, Germany}
\altaffiltext{6}{Heidelberg Institut f\"ur Theoretische Studien,
  Schloss-Wolfsbrunnenweg 35, 69118 Heidelberg, Germany}

\email{jwise@physics.gatech.edu}

\begin{abstract}

By definition, Population III stars are metal-free, and their
protostellar collapse is driven by molecular hydrogen cooling in the
gas-phase, leading to large characteristic masses.  Population II
stars with lower characteristic masses form when the star-forming gas
reaches a critical metallicity of $10^{-6} - 10^{-3.5}~Z_\odot$.  We
present an adaptive mesh refinement radiation hydrodynamics simulation
that follows the transition from Population III to II star formation.
The maximum spatial resolution of 1 comoving parsec allows for
individual molecular clouds to be well-resolved and their stellar
associations to be studied in detail.  We model stellar radiative
feedback with adaptive ray tracing.  A top-heavy initial mass function
for the Population III stars is considered, resulting in a plausible
distribution of pair-instability supernovae and associated metal
enrichment.  We find that the gas fraction recovers from 5 percent to
nearly the cosmic fraction in halos with   merger histories rich in
halos above $10^7 \Ms$.  A single pair-instability supernova is
sufficient to enrich the host halo to a metallicity floor of $10^{-3}
Z_\odot$ and to transition to Population II star formation.  This
provides a natural explanation for the observed floor on damped Lyman
alpha (DLA) systems metallicities reported in the literature, which is
of this order.  We find that stellar metallicities do not necessarily
trace stellar ages, as mergers of halos with established stellar
populations can create superpositions of $t-Z$ evolutionary tracks.  A
bimodal metallicity distribution is created after a starburst occurs
when the halo can cool efficiently through atomic line cooling.

\end{abstract}


\section{Motivation}

The first (Pop III) stars are metal-free and have a large
characteristic mass.  Previously, these stars were believed to be
massive, with universally suppressed fragmentation \citep{ABN02,
Bromm02_P3, OShea07a}.  Recently, several studies have suggested that
while fragmentation at times prior to the formation of the first
protostar may occur at scales of $1000-2000~\mathrm{AU}$, such
fragmentation is likely to result in high-mass, low-multiplicity
groups \citep{2009Sci...325..601T, Stacy10_Binary}.  Subsequent
studies have suggested that at later times, subsequent to the
formation of the first protostar, additional fragmentation may proceed
\citep{Greif11_P3Cluster, Clark11_Frag}.  A fraction of high-mass,
metal-free stars enrich the surrounding intergalactic medium (IGM)
when they go supernova, which can happen in stars $\lsim 40~\Ms$ in
Type II supernovae (SNe) or in stars roughly between 140 \Ms~and 260
\Ms~in pair-instability SNe \citep[PISNe;][]{2002ApJ...567..532H}.
The host halo and the neighboring halos are then enriched with this
ejecta.  There exists a critical metallicity that is $\sim 10^{-6}
Z_\odot$ if dust cooling is efficient \citep{Omukai05,
Schneider06_Frag, clark08} and $\sim 10^{-3.5} Z_\odot$ otherwise
\citep{Bromm01, 2009ApJ...691..441S}, where the gas can cool rapidly,
lowering its Jeans mass.  An intermediate characteristic mass of
$\sim10~\Ms$ can occur if the gas cooling is suppressed to the cosmic
microwave background (CMB) temperature \citep{Larson98,
Tumlinson07_IMF, 2009ApJ...691..441S}.  The resulting Population II
star cluster will thus have a lower characteristic stellar mass than
its metal-free progenitors.  These first stellar clusters may be
connected to stars in the Milky Way halo and nearby dwarf spheroidal
(dSph) galaxies, both with a metallicity floor of [Z/H]\footnote{We
use the conventional notation, [Z/H] $\equiv$ log(Z/H) -
log(Z$_\odot$/H$_\odot$).} = --4 \citep{Beers05, Frebel10_Obs,
Tafelmeyer10}.

Pop III stars primarily form in dark matter halos with $M \sim 10^6
\Ms$.  Their stellar radiation ionizes and heats an \ion{H}{2} region
with a radius 1--3 kpc.  The over-pressurized \ion{H}{2} region drives
a 30 \kms~shock that is 10 times greater than the escape velocity of
the halo, leaving behind a warm ($3 \times 10^4$ K) and diffuse (0.5
\cubecm) medium \citep{Kitayama04, Whalen04, Abel07}.  This aids in
the expansion of the blastwave because it delays the transition to the
Sedov-Taylor and snowplow phases.  In PISNe, approximately half of the
metals stay in the IGM with a metal bubble size of a few kpc
\citep{Wise08_Gal, Greif10}.  The blastwave may induce star formation
(SF) in nearby halos through the compression of the gas
\citep{Ferrara98}, and timescales for metal mixing into the dense gas
are many dynamical times \citep{Cen08} for shock velocities
$\lsim100~\kms$.

The transition from Pop III to Pop II SF is solely dependent on the
propagation of metals from the SNe remnants into future sites of SF.
Their flows are complex because of the interactions between the SN
blastwave, cosmological accretion, halo mergers, and nearby stellar
feedback.  Spurred by the notion of a critical metallicity, this
transition has been extensively studied with (i) volume-averaged
semi-analytic models \citep{Scannapieco03, Yoshida04,
  Furlanetto05_Reion}, (ii) models using hierarchical merger trees
\citep{Tumlinson06, Tumlinson10, Salvadori07, Komiya10}, (iii)
post-processing of cosmological simulations with blastwave models
\citep{Karlsson08, Trenti09, Trenti10}, and (iv) direct numerical
simulations with stellar feedback \citep{Tornatore07, Ricotti08,
  Maio10_Pop32, Maio11_Enrich}.

Over the past decade, these works have refined the general picture of
inhomogeneous metal enrichment and the transition to Pop II star
formation, and here we give an overview of its development.
Considering only Pop III star formation, \citeauthor{Yoshida04}
calculated that Pop III stars can raise the mean metallicity to $-4.5
\lsim$ [Z/H] $\lsim -3.5$ by redshift 15 in the upper limit where all
metal-free stars have $M = 200 \Ms$ and produce PISNe.  Considering
both Pop III and II star formation, \citeauthor{Scannapieco03} found
that the transition between the two modes is a gradual process where
both modes are coeval, confirmed by most of the later works.  Because
the host galaxies have small masses and are subject to negative
radiative feedback through photo-heating and photo-dissociation, the
minimum halo mass gradually increases with the radiation background
intensity, which \citeauthor{Furlanetto05_Reion} found to delay metal
enrichment and place it closer to the epoch of reionization.
\citeauthor{Trenti09} noted that underdense regions of the universe
are still pristine at $z=6$ with Pop III stars still forming at these
late epochs.  This group later expanded on these results to find that
Pop II SFR becomes dominant at $z<25$ in the buildup of a Milky Way
(MW) like halo.  Furthermore they stress the importance of a
photo-dissociating radiation background that reduces enrichment by
PISNe and increases the importance of metal-rich galactic outflows,
where only $10^{-4} - 10^{-2}$ of PISN ejecta ends up in extremely
metal poor (EMP) stars with $M > 0.9 \Ms$.

The Pop III IMF should play an important role in abundance patterns in
EMP stars in the MW halo and the physics of reionization.  Using these
data as constraints, \citeauthor{Tumlinson06} found that Pop III IMFs
with log-normal distributions with mean masses between 8 and 42
\Ms~best fit the data.  Furthermore he concludes that EMP stars with
[Z/H] $<$ --3 have between 1 and 10 metal-free SN progenitors and
the Pop III SFR is less than 1\% of the total SFR at $z=6$.
\citeauthor{Karlsson08} use observational data of EMP stars with
their model to constrain the mass fraction of Pop III stars that
die with a PISN is less than 40 per cent.  They also conclude that
stars enriched primarily by PISNe have [Ca/H] $\gsim$ --2.6, which
could have been missed by some EMP surveys.

These works have clearly shown that many processes play a key role in
the transition from Pop III to II stars, but how does their interplay
affect the formation of the first galaxies?  Will metal-free pockets
of gas still exist in dwarf galaxies that are already forming
metal-enriched stars?  How do the metal-enriched ejecta propagate into
future sites of star formation?  Numerical simulations are powerful
tools to detangle and study these complexities.  To answer such
questions, we have carried out a simulation that includes both types
of SF and their radiative and mechanical feedback.  The methods used
here incorporate and link together recent results from metal-enriched
and metal-free star formation, the critical metallicity, and
pair-instability supernovae.  This is the first time it has been
possible to correlate the formation and feedback of the first stars to
protogalaxies, resolving individual molecular clouds and minihalos
down to $10^5 \Ms$ and including the most important physical effects,
such as radiative and supernova feedback.

We first describe the simulation setup and numerical methods behind
our work in \S\ref{sec:setup}.  Then in \S\ref{sec:results} we present
the evolution of two selected high-redshift dwarf galaxies, focusing
on their metallicities and stellar populations.  Last we discuss
possible dependencies of our results on the Pop III initial mass
function and conclude in \S\ref{sec:discuss}.

\section{Method}
\label{sec:setup}

\subsection{Simulation setup}

We use the adaptive mesh refinement (AMR) code
\enzo~v2.0\footnote{\texttt{enzo.googlecode.com, changeset
    b86d8ba026d6}} \citep{OShea2004}.  It uses an $N$-body adaptive
particle-mesh solver \citep{Efstathiou85, Couchman91, BryanNorman1997}
to follow the dark matter (DM) dynamics.  It solves the hydrodynamical
equation using the second-order accurate piecewise parabolic method
\citep{Woodward84, Bryan95}, while a Riemann solver ensures accurate
shock capturing with minimal viscosity.  We use the recently added
HLLC Riemann solver \citep{Toro94_HLLC} for additional stability in
strong shocks and rarefaction waves.  We use the nine-species
(\ion{H}{1}, \ion{H}{2}, \ion{He}{1}, \ion{He}{2}, \ion{He}{3}, e$^-$,
\hh, \hh$^+$, H$^-$) non-equilibrium chemistry model in
\enzo~\citep{Abel97, Anninos97} and the \hh~cooling rates from
\citet{Glover08_Rates}.  We spatially distinguish metal enrichment
from Pop II and Pop III stars.  We will follow-up this study with one
that considers radiative cooling from metal-enriched gas, using the
method of \citet{2008MNRAS.385.1443S} that is already implemented in
\enzo.

To resolve minihalos with at least 100 dark matter (DM) particles and
follow the formation of the first generation of dwarf galaxies, we use
a simulation box of 1 Mpc that has a resolution of $256^3$.  This
gives us a DM mass resolution of 1840 \Ms.  We refine the grid on
baryon overdensities of $3 \times 2^{-0.2l}$, where $l$ is the AMR
level, resulting in a super-Lagrangian behavior \citep[also
  see][]{OShea08}.  We also refine on a DM overdensity of three and
always resolve the local Jeans length by at least 4 cells, avoiding
artificial fragmentation during gaseous collapses \citep{Truelove97}.
If any of these criteria are met in a single cell, it is flagged for
further spatial refinement.

We initialize the simulation with grafic \citep{Bertschinger01} at $z
= 130$ and use the cosmological parameters from the 7-year WMAP
$\Lambda$CDM+SZ+LENS best fit \citep{WMAP7}: $\Omega_M = 0.266$,
$\Omega_\Lambda = 0.734$, $\Omega_b = 0.0449$, $h = 0.71$, $\sigma_8 =
0.81$, and $n = 0.963$ with the variables having their usual
definitions.  We use a maximum refinement level of $l = 12$, resulting
in a maximal comoving resolution of 1 pc.  The adaptive particle-mesh
solver has a force resolution of two cell widths of a given AMR grid.
We stop the simulation at $z=7$ to prevent any large-scale modes with
$r \sim L_{\rm box}/2$ from entering the non-linear growth regime.  At
the final redshift, the simulation has $1.4 \times 10^8$ computational
cells and required 250,000 CPU hours, using 512 cores.

\subsection{Star formation}

We distinguish Pop II and Pop III SF by the total metallicity of the
densest cell in the molecular cloud.  Pop II stars are formed if [Z/H]
$> -4$, and Pop III stars are formed otherwise.  We do not consider
Pop III.2 stars and intermediate mass stars from CMB-limited cooling.

\subsubsection{Population III star formation}

Simulations following the full accretion history of Pop III stars,
from their formation inside cosmological minihalos through
transitioning onto the main sequence, are not currently possible.
However, under the assumption of spherical accretion, reasonable
estimates of the mass ranges for Pop III stars can be made by
extrapolating forward in time the instantaneous accretion rate at the
time of formation of the first protostellar core and comparing the
timescales for accretion against the Kelvin-Helmholtz time.  Several
studies, conducting this extrapolation, have shown that the
characteristic mass of Pop III stars $M_{\rm char} \sim 100~\Ms$
\citep{ABN02, Bromm02_P3, OShea07a, Yoshida08}.  They form in
molecular clouds that coexist with the dark matter halo center with
limited fragmentation occurring during their collapse; however
\citet{2009Sci...325..601T} and \citet{Stacy10_Binary} have shown that
Pop III binaries may form in a fraction of such halos.  There have
been some recent hints that the accretion disk around the first
protostar may fragment and produce clustered Pop III star formation
\citep{Clark11_Frag, Greif11_P3Cluster}.  The final stellar masses in
the aforementioned calculations are still highly uncertain with a
limited number of samples.  Nevertheless, they all point toward an IMF
with a larger characteristic mass than present-day star formation.


For Pop III stars, we use the same SF model as \citet{Wise08_Gal}
where each star particle represents a single star.  In this model, a
star particle forms when a cell has all of the following criteria:
\begin{enumerate}
\item an overdensity of $5 \times 10^5$ ($\sim 10^3\; \cubecm$ at
  $z=10$),
\item a converging gas flow ($\nabla \cdot \mathbf{v}_{\rm gas} < 0$),
  and
\item a molecular hydrogen fraction $f_{\rm H2} > 5 \times 10^{-4}$.
\end{enumerate}
These physical conditions are typical of collapsing metal-free
molecular clouds $\sim 10$ Myr before the birth of a Pop III
main-sequence star \citep{ABN02, OShea07a}.  This prescription is
similar to the \citet{Cen92} method but removes the Jeans unstable
($M_{\rm gas} > M_J$) and cooling timescale ($t_{\rm cool} < t_{\rm
  dyn}$) requirements.  We do not consider the former criterion
because it is not applicable to simulations that resolve the Jeans
length at all times.  The molecular hydrogen fraction requirement
effectively constrains star formation to cooling molecular clouds,
where the \hh~formation rate is significantly larger than the
dissociation rate from a Lyman-Werner radiation field.

If multiple cells meet the star particle formation criteria within 1
pc, we form one Pop III star particle with the center of mass of these
flagged cells to ensure that one massive star is created per
metal-free molecular cloud.  Instead of using a fixed stellar mass
like in our previous work, we randomly sample from an IMF with a
functional form of
\begin{equation}
\label{eqn:imf}
f(\log M) dM = M^{-1.3} \exp\left[-\left(\frac{M_{\rm
      char}}{M}\right)^{1.6}\right] dM
\end{equation}
to determine the stellar mass.  Above $M_{\rm char}$, it behaves as a
Salpeter IMF but is exponentially cutoff below that mass
\citep{Chabrier03, Clark09}.  For reproducibility, we record the
number of times the random number generator (Mersenne twister;
\citet{MTwister}) has been called for use when restarting the
simulations.  We stress that the Pop III IMF is still highly
uncertain.  In Sec. \ref{sec:discuss}, we discuss possible
dependencies of our results to changes in the IMF.  

After the star particle forms and its mass is randomly sampled from
the IMF, an equal amount of gas is removed from the computational grid
in a sphere that contains twice the stellar mass and is centered on
the star particle.  The star particle acquires the mass-weighted
velocity of the gas contained in this sphere.

%

\subsubsection{Population II star formation}

We treat Pop II star formation with the same prescription as
\citet{Wise09}, which is similar to the Pop III prescription but
without the minimum \hh~fraction requirement.  This is removed because
the metal-enriched gas can efficiently cool even in the presence of a
strong UV radiation field \citep[e.g.][]{Safranek10}.  To ensure the
volume is cooling, we restrict star formation to gas with temperatures
$T < 1000$ K.  Unlike Pop III star particles that represent individual
stars, Pop II star particles represent a star cluster of some total
mass and an assumed normal IMF.

Once a cell meets these criteria, the prescription searches outward
with increasing radius for the boundary of the molecular cloud,
centered on the most dense cell.  Here, we define a molecular cloud as
a sphere with a dynamical time $t_{\rm dyn} = 3$ Myr (corresponding to
an average density $\bar{\rho}_{\rm MC} \simeq 1000\mu\; \cubecm$) and
a radius $R_{\rm MC}$, where $\mu$ is the mean molecular weight.  Once
the sphere radius is found, a fraction $c_\star = 0.07 f_{\rm cold}$
of the cold gas ($T < 10^3$ K) inside the sphere is converted into a
star particle with mass $m_\star = c_\star (4\pi/3) \bar{\rho}_{\rm
  MC} R_{\rm MC}^3$.  This treatment of cold gas accretion is similar
to a star formation model with a multi-phase subgrid model
\citep{Springel03b} but is employed in a simulation that can resolve
the multi-phase interstellar stellar medium (ISM).

After the star particle is created, we replace the sphere with a
uniform density $\rho_{\rm MC} = (1 - c_\star) / (Gt_{\rm dyn}^2)$ and
temperature $T = 10^4$ K, which approximates the initial stages of an
\ion{H}{2} region.  Our choice of $c_\star$ agrees with the theory
proposed in \citet{Krumholz05} that are supported by observations.
Observations of star-forming regions show that the dynamical time is
approximately 700 kyr and that SF is extended over several dynamical
times \citep[e.g.][]{Tan06}, which motivates our choice of $t_{\rm
  dyn}$ of our definition of molecular clouds in our SF prescription.

Similar to the Pop III treatment, we merge any newly created particles
within a radius of 10 pc.  The new star particle acquires the center
of mass of the merged particles.  During the merging, the gas is
unaltered besides the initial gas accretion from the grid to the star
particles.  We set the minimum mass of a star particle to $m_{\rm
  \star, min} = 1000 \Ms$.  If the initial mass does not exceed
$m_{\rm \star, min}$, the star particle does not provide any feedback
and continues to accrete until it reaches $m_{\rm \star, min}$.

\subsection{Stellar feedback}

The radiation field is evolved with adaptive ray tracing
\citep{Abel02_RT, Wise11_Moray} that is based on the HEALPix framework
\citep{HEALPix} and is coupled self-consistently to the hydrodynamics.
In this work, we only consider energy coupling to the gas; we defer an
investigation of the effects of radiation momentum coupling to a
subsequent paper.  Each star particle is a point source of hydrogen
ionizing radiation with the ionizing luminosity equally split between
48 initial rays (HEALPix level 2).  We use a mono-chromatic spectrum
for the radiation with the energy $E_{\rm ph}$ equaling the
luminosity-weighted photon energy of the spectrum.  For a cosmological
simulation that focuses on galaxies, this does not significantly
affect the overall galactic dynamics \citep[see Sec. 6.3
  in][]{Wise11_Moray}.  We do not consider helium singly- and
doubly-ionizing radiation.  As the rays propagate from the source or
into a high resolution AMR grid, they are adaptively split into 4
child rays, increasing the angular resolution of the solution, when
the solid angle of the ray $\theta = 4\pi/(12 \times 4^{L})$ is larger
than 1/3 of the cell area, where $L$ is the HEALPix level.  We model
the \hh~dissociating radiation with an optically-thin, inverse square
profile, centered on all Pop II and III star particles.  In principle,
self-shielding effects can be included by utilizing our ray tracing
framework or other approximate methods that calculate the \hh~column
density using the local Sobolev length \citep{Wolcott11} or only
integrating along Cartesian axes or random lines of sight
\citep{Yoshida03, Yoshida07_HII, Glover07}.  In halos with $M \ga 10^8
\Ms$, the \hh~column density may become large enough to self-shield
itself from LW radiation \citep{Wise08_Gal}.  Thus we may be
underestimating the SFRs in our work.

\subsubsection{Population III stellar feedback}

We use mass-dependent hydrogen ionizing and LW photon luminosities and
lifetimes of the Pop III stars from \citet{Schaerer02}.  We use a
mass-independent photon energy $E_{\rm ph} = 29.6$ eV, appropriate for
the near-constant $10^5$ K surface temperatures of Pop III stars.
They die as Type II SNe if $11 \le M_\star/\Ms \le 40$
\Ms~\citep{Woosley95} and as PISNe if they are in the mass range
140--260 \Ms~\citep{Heger03}, where $M_\star$ is the stellar mass.
For normal Type II SNe between 11--20 \Ms, we use an explosion energy
of \tento{51}~erg and a linear fit to the metal ejecta mass calculated
in \citet{Nomoto06},
\begin{equation}
  \label{eqn:typeii}
  M_{\rm Z}/\Ms = 0.1077 + 0.3383 \times (M_\star/\Ms - 11).
\end{equation}
We model the SNe of stars with $20 \le M_\star/\Ms \le 40$ as
hypernova with the energies and ejecta masses also taken from
\citeauthor{Nomoto06}, linearly interpolating their results to
$M_\star$.  For PISNe, we use the explosion energy from
\citet{Heger02}, where we fit the following function to their models,
\begin{equation}
  \label{eqn:pisn}
  E_{\rm PISN} = 10^{51} \times [5.0 + 1.304 (M_{\rm He}/\Ms - 64)] \;
  \mathrm{erg},
\end{equation}
where $M_{\rm He} = (13/24) \times (M_\star - 20) \Ms$ is the helium
core (and equivalently the metal ejecta) mass and $M_\star$ is the
stellar mass.  If the stellar mass is outside of these ranges, then an
inert, collisionless black hole (BH) particle is created.  The
blastwave is modeled by injecting the explosion energy and ejecta mass
into a sphere of 10 pc, smoothed at its surface to improve numerical
stability \citep{Wise08_Gal}.  Because we resolve the blastwave
relatively well with several cells across at its initialization, the
thermal energy is converted into kinetic energy and agrees with the
Sedov-Taylor solution \citep[e.g.][]{Greif07}.

\subsubsection{Population II star formation}

The Pop II stars emit 6000 hydrogen ionizing photons per baryon over
their lifetime and $E_{\rm ph} = 21.6$ eV, appropriate for a [Z/H] =
$-1.3$ population.  We note that lower-metallicity stars generate up
to 60\% more ionizing photons and might be underestimating the
radiative feedback \citep{Schaerer03}.  Nonetheless this study
provides an excellent first insight in the transition to Pop II SF, as
the metal enrichment is the key ingredient here.  The star particles
live for 20 Myr, the maximum lifetime of an OB star.  These stars
generate the majority of the ionizing radiation and SNe feedback in
stellar clusters, thus we ignore any feedback from lower mass stars.

For SN feedback, these stars generate $6.8 \times 10^{48}$ erg
s$^{-1}$ $\Ms^{-1}$ after living for 4 Myr, which is injected into
spheres with a radius of 10 pc.  However if the resolution of the grid
is less than 10/3 pc, we distribute the energy into a $3^3$ cube
surrounding the star particle.  In the same region, star particles
also return ejected material with mass
\begin{equation}
  \label{eqn:SNmass}
  \Delta m_{\rm ej} = \frac{0.25 \; \Delta t \; M_\star} 
         {t_0 - 4\;\mathrm{Myr}}
\end{equation}
back to the grid at every timestep on the finest AMR level.  Here
$t_0$ = 20 Myr is the lifetime of the star particle.  This ejected gas
has solar metallicity $Z = 0.02$, resulting in a total metal yield $y
= 0.005$.  In addition to numerical stability, energy distribution
across several cells instead of a single cell has been shown to match
cosmic SFRs better in galaxy simulations at lower redshifts
\citep{Smith11}.

\section{Results}
\label{sec:results}


\begin{figure*}
\epsscale{1.15}
\plotone{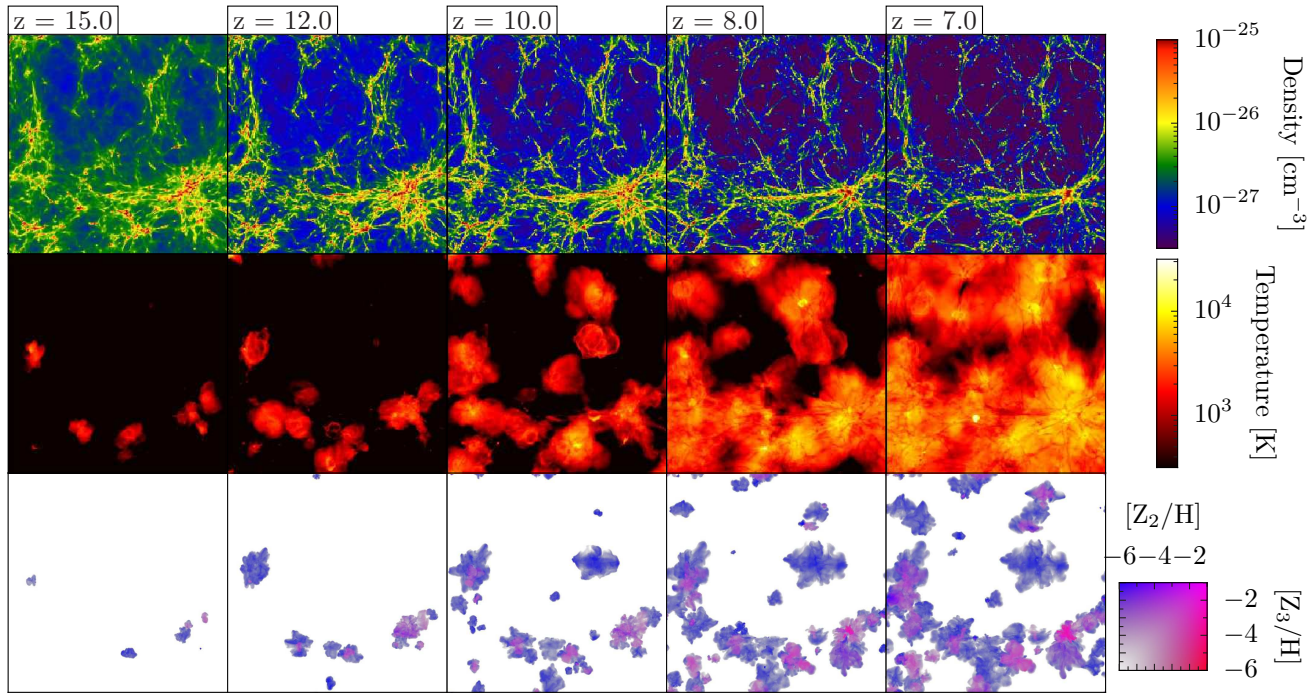}
\epsscale{1}
  \caption{\label{fig:evo-mosaic} Evolution of the entire simulation
    volume ($L_{\rm box} = 1$ Mpc) at redshifts 15, 12, 10, 8, and 7
    (left to right).  Pictured here are the density-weighted
    projections of density (top), temperature (middle), and
    metallicity (bottom).  Note how the stellar radiative feedback
    from low-mass galaxies reionize the majority of the volume.  The
    metallicity projections are a composite image of metals
    originating from Pop II (red) and III (blue) stars with magenta
    indicating a mixture of the two.}
\end{figure*}



\begin{figure}
\epsscale{1.15}
\plotone{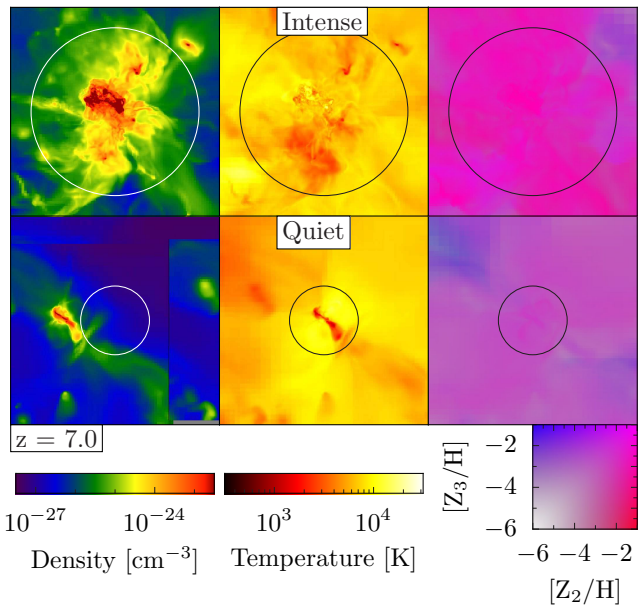}
\epsscale{1}
  \caption{\label{fig:halo-mosaic} Density-weighted projections of
    density (left), temperature (center), and metallicity (right) of
    the two selected halos at $z=7$.  The field of view is 5 proper
    kpc, and the circles have a radius of $r_{200}$.  The composite
    metallicity images are constructed in the same fashion as
    Fig. \ref{fig:evo-mosaic}.}
\end{figure}


\begin{deluxetable}{ccccc}
\tablecolumns{5}
\tabletypesize{}
\tablewidth{\columnwidth}
\tablecaption{Enrichment Fractions at Redshift 7\label{tab:enrich}}

\tablehead{\colhead{[Z/H]} & 
  \colhead{$f_v$(Pop III)} & \colhead{$f_m$(Pop III)} & 
  \colhead{$f_v$(Pop II)} & \colhead{$f_m$(Pop II)}
}
\startdata
-6 & 3.81(-2) & 1.43(-1) & 5.52(-3) & 5.22(-2) \\
-5 & 3.28(-2) & 1.24(-1) & 3.92(-3) & 4.36(-2) \\
-4 & 2.65(-2) & 1.01(-1) & 2.33(-3) & 3.43(-2) \\
-3 & 1.86(-2) & 6.45(-2) & 1.00(-3) & 2.50(-2) \\
-2 & 8.29(-3) & 1.23(-2) & 2.61(-4) & 1.73(-2)
\enddata

\tablecomments{$f_v$ and $f_m$ are volume filling and mass-weighted
  fractions, respectively, above several different metallicity cuts.
  Values are listed in scientific notation with the exponents in
  parentheses.}

\end{deluxetable}

Here we first present the global star formation rate (SFR) during
early galaxy formation and then focus on the gaseous and stellar
evolution of two selected halos in the simulation: one that has an
early mass buildup but no major mergers after $z=12$, and one that
experiences a series of major mergers between $z=10$ and $z=7$.  We
name the halos ``quiet'' and ``intense'', respectively.  The entire
simulation contains 38 galaxies with 3640 Pop II stellar clusters and
captures the formation of 333 Pop III stars.

We illustrate the evolution of the simulation from $z=15$ to $z=7$ in
Figure \ref{fig:evo-mosaic} with density weighted projections of gas
density, temperature, and metallicity of the entire box.  Figure
\ref{fig:halo-mosaic} shows the same types of plots but focusing on
the two halos of interest with a field of view of 5 proper kpc.
Radiative and mechanical feedback create a multi-phase medium inside
these halos, which are embedded in a warm and ionized IGM.  At the
final redshift, the volume and mass ionized fraction are 0.76 and
0.72, respectively, which are consistent with large-scale reionization
simulations \citep{Zahn11}.  PISNe enrich 0.065 and 0.019 of the mass
and volume, respectively, above $10^{-3} \zsun$, and 0.025 and 0.0010
of the mass and volume are enriched by Type II SNe from Pop II stars
above the same threshold.  With $M_{\rm char} = 100 \Ms$, a negligible
fraction of Pop III stars die as Type II SNe; thus, all metals
originating from Pop III stars are produced in PISNe.  Table
\ref{tab:enrich} gives additional enriched fractions with different
metallicity cuts.

\subsection{Global star formation rate}


\begin{figure}
\epsscale{1.15}
\plotone{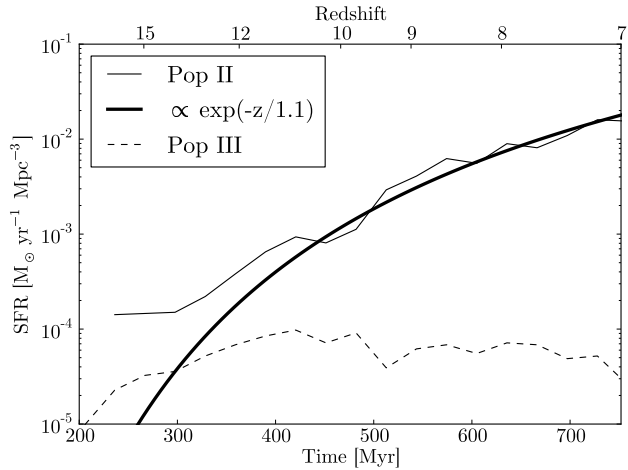}
\epsscale{1}
\caption{\label{fig:sfr} Cosmic star formation rates for Pop II
  (solid) and III (dashed).  Note that the Pop III rates are nearly
  constant from redshift 12 to 7.  Pristine halos still survive in
  underdense regions, supporting late-time metal-free star formation.}
\end{figure}


Figure \ref{fig:sfr} shows the cosmic SFR in the simulation for Pop II
and III stars.  Metal-free stars form at a nearly constant rate of $5
\times 10^{-5}$ \sfr~until the end of the simulation.  The
dissociating and ionization radiation from nearby Pop III stars and
galaxies do not halt the formation of such stars in low-mass halos
but only suppress them.  We expect a uniform \hh~dissociating
radiation background to further delay them \citep{Machacek01, Wise07,
  OShea08}.

The cosmic Pop II SFR is best fit with an exponential $\exp(-z/1.1)$.
Metal-enriched star formation begins early in the simulation at $z
\sim 16$ when a PISN blastwave overtakes a satellite halo with $M \sim
5 \times 10^5 \Ms$, triggering star formation there.  At $z>10$, these
enriched minihalos cause the Pop II SFR to be higher than this
relation.  Only after $z \sim 10$ does the SFR follows this
exponential, which is a steeper time dependency than the $\exp(-z/3)$
relation found in \citet{Hernquist03}, who also acknowledged that this
behavior should occur at early times.  At redshift 7, it reaches $2
\times 10^{-2}$ \sfr.  This value agrees with derived SFRs from
$z=6-10$ galaxy luminosity functions \citep[e.g.][]{Bouwens11} and
large-scale galaxy formation simulations \citep[e.g.][]{Schaye10}.

\subsection{Mass evolution}
\label{sec:halo}



\begin{figure*}
\epsscale{1.15}
\plotone{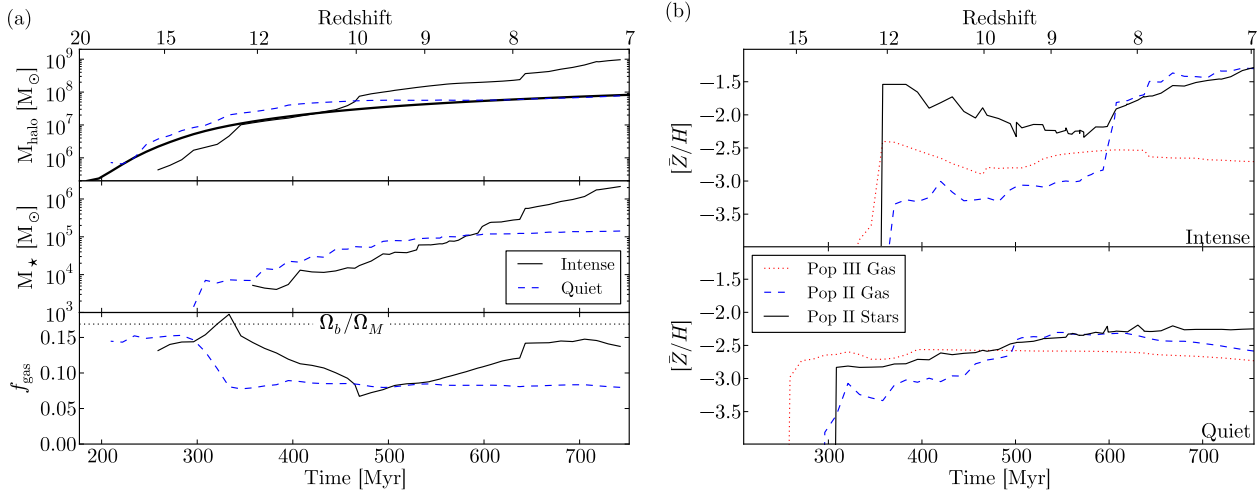}
\epsscale{1}
\caption{\label{fig:evo} (a) Evolution of the total halo mass (top),
  stellar mass (middle), and gas fraction (bottom) of the quiet
  (dashed) and intense (solid) halos.  In the top panel, the filtering
  halo mass is plotted as the thick black line (overlapping with the
  quiet halo after $z=9$.  (b) Mass-weighted stellar metallicities and
  gas metallicities enriched by Pop II and Pop III SNe of the intense
  (top) and quiet (bottom) halos.}
\end{figure*}


Figure \ref{fig:evo}a shows the total, metal-enriched stellar, and gas
mass history of the most massive progenitors of both halos.  The quiet
halo undergoes a series of major mergers at $z > 12$, growing by a
factor of 30 to $2.5 \times 10^7 \Ms$ within 150 Myr.  Subsequently,
it only grows by a factor of 3 by $z=7$ mainly through smooth
accretion from the filaments and IGM.  It is the most massive halo in
the simulation between redshifts 13 and 10.  At the same time, the
intense halo has a mass $M = 3 \times 10^6 \Ms$, but it is contained
in a biased region on a comoving scale of 50 kpc with $\sim25$ halos
with $M \sim 10^6 \Ms$.  After $z=10$, these halos hierarchically
merge to form a $10^9 \Ms$ halo at $z=7$ with two major mergers at
redshifts 10 and 7.9, seen in the rapid increases in total mass in
Figure \ref{fig:evo}a.  The merger history of the two halos are not
atypical as dark matter halos can experience both quiescent and
vigorous mass accretion rates.

Both halos start forming Pop II stars when $M = 10^7 \Ms$.  This is
consistent with the filtering mass $M_f$, which is plotted as a thick
line in Figure \ref{fig:evo}a, of high-redshift halos when it accretes
mainly from a pre-heated IGM \citep{gnedin98, gnedin00, Wise08_Gal}.
Afterwards these halos can cool efficiently through \hh~cooling,
sustaining constant and sometimes bursting SF.  The latter
characteristics are equated with the definition of a galaxy.  The
quiet halo forms $10^5 \Ms$ of stars by $z=9$.  This initial starburst
photo-evaporates the majority of its molecular clouds, in addition to
heating and ionizing the surrounding IGM out to a radius of 10--15 kpc
at $z=9$.  These respectively reduce the in-situ and external cold gas
supply that could feed future SF.

The gas fractions of both halos decrease from 0.15 to 0.08 by outflows
driven by ionization fronts and blastwaves in their initial
starbursts.  The quiet halo does not have a major merger with any halo
with $M > M_f$, leading to a small final gas fraction.  These low-mass
halos are photo-evaporated, hosting diffuse warm gas reservoirs
instead of cold dense cores.  After $z=10$, the halo mainly accretes
warm diffuse gas from the filaments and IGM.  In contrast, the intense
halo grows from major mergers of halos with $M > M_f$.  The progenitor
halos involved in the major mergers are able to host molecular clouds
and have higher gas fractions.  Between $z=10$ and $z=8$, the gas
fraction increases from 0.07 to 0.12 until it jumps to 0.14 when a
gas-rich major merger occurs.  The stellar mass accordingly increases
with the ample supply of gas during this period.  

\subsection{Metallicity evolution}

The evolution of the stellar and gas metallicity of both halos are
illustrated in Figure \ref{fig:evo}b.  PISNe from Pop III stars enrich
the nearby IGM out to a radius of 10 kpc and provides a metallicity
floor of $[Z_3/H] \sim -3$.  Ejecta from Pop II SNe initially enrich
the ISM of both halos to an average $[\bar{Z}_2/H]$ between --3.5 and
--3.

In the quiet halo, an equilibrium of $[\bar{Z}_2/H] \sim -2.5$ is
established between metal-rich outflows and metal-poor inflows.
Galactic outflows are directed in the polar directions of the gas
disk, keeping the adjacent filaments metal-poor.  These features and a
well-mixed ISM \citep[cf.][]{Wise08_Gal, Greif10} are apparent in the
metallicity projections in Figure \ref{fig:halo-mosaic}.  The average
stellar metallicity is within 0.5 dex of $[\bar{Z}_2/H]$.

In the intense halo, the first few Pop II star clusters have $[Z/H]$
between --1 and --2 and dominate the average stellar metallicity at $z
> 8$.  The metallicity later increases by a factor of 30 to
$[\bar{Z}_2/H] = -1.5$ through self-enrichment from a starburst.
Because this halo is located in a large-scale overdensity, most of the
ejecta falls back into the halo after reaching distances up to 20
comoving kpc, keeping the halo gas metallicity high because the
inflows are relatively metal-rich themselves.  After the $z \sim 8$
starburst, the average stellar metallicity follows the average gas
metallicity within 0.1 dex.


\begin{figure}
\epsscale{1.15}
\plotone{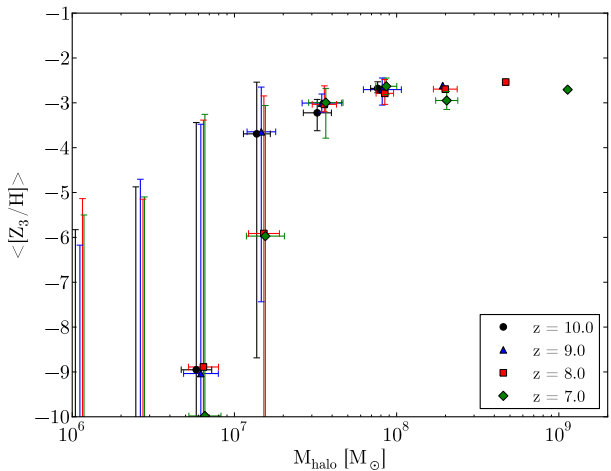}
\epsscale{1}
\caption{\label{fig:z3} Average halo metallicity from Pop III SNe as a
  function of total halo mass.  Inhomogeneous metal enrichment causes
  the large spread in metallicity at low masses.  At $M_{\rm halo} > 3
  \times 10^7 \Ms$, all halos are enriched by Pop III SNe above [Z/H]
  $>$ --4.  The y-axis error bars show the 25\% and 75\% percentile
  values because the underlying distribution is not Gaussian.}
\end{figure}



\begin{figure*}
\plottwo{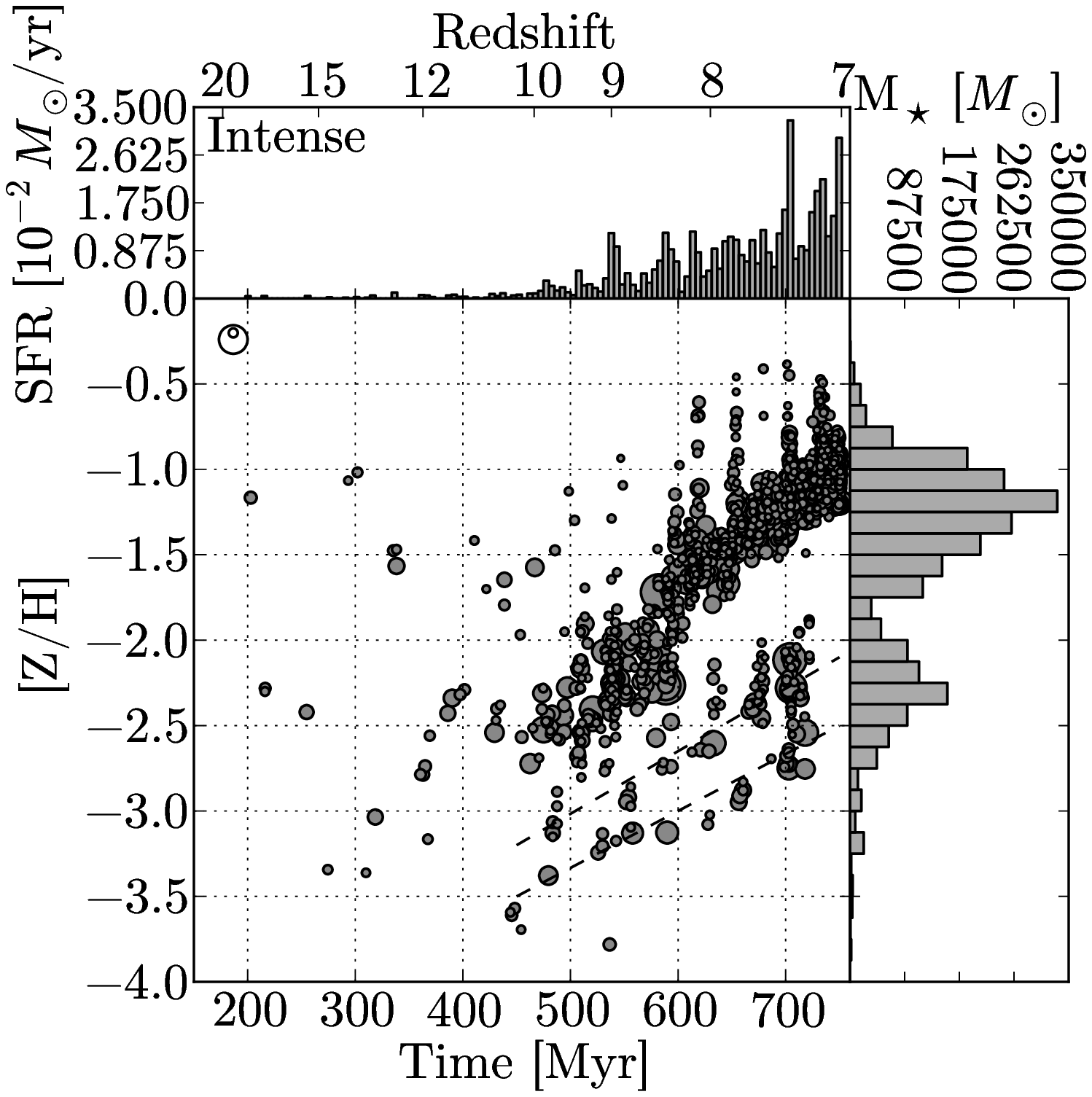}{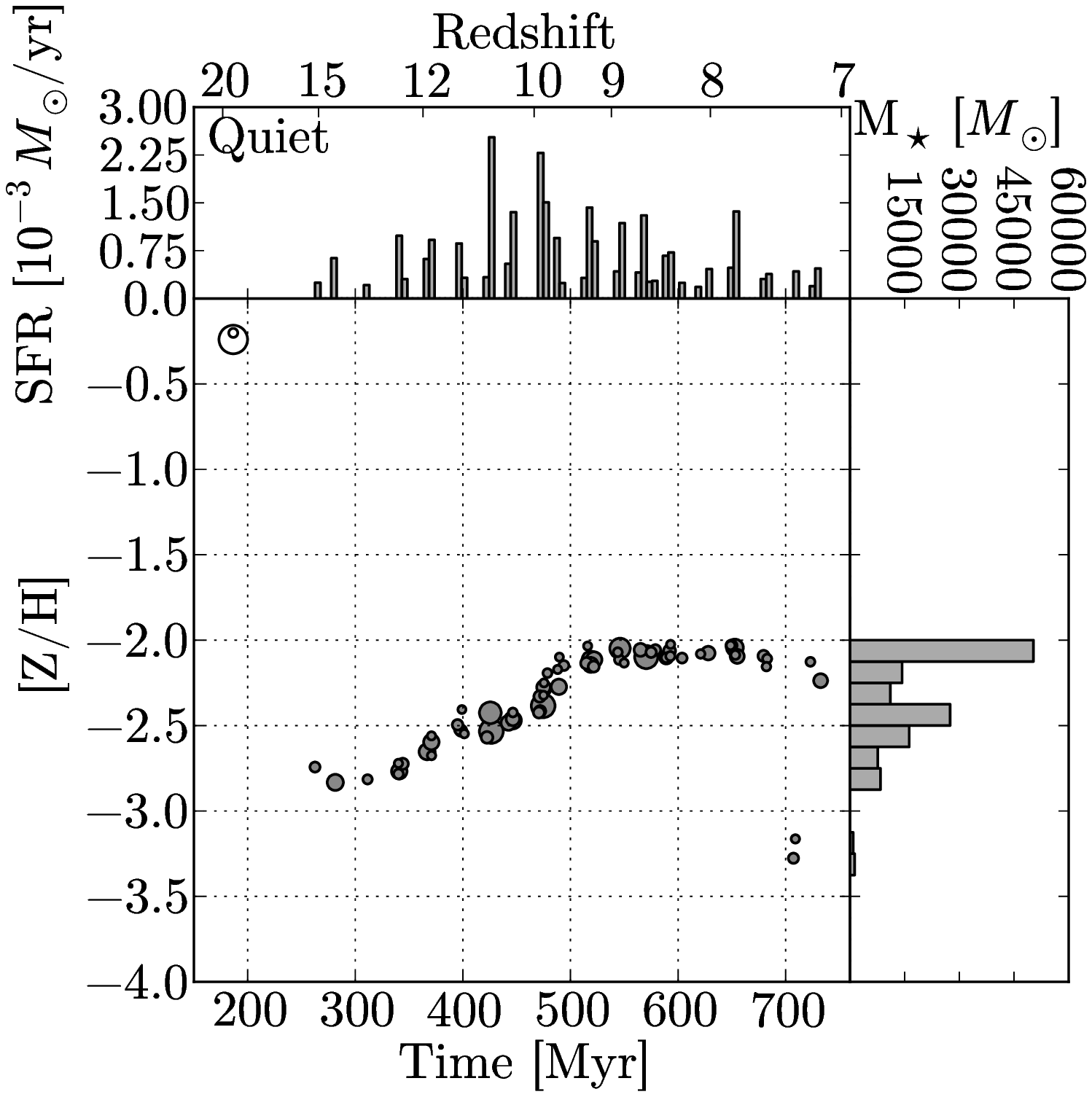}
\caption{\label{fig:pops} The scatter plots show the metal-enriched
  (Pop II) star formation history of the intense (left) and quiet
  (right) halos as a function of total metallicity, i.e. the sum of
  metal ejecta from both Pop II and Pop III SNe, at $z=7$.  Each
  circle represents a star cluster, whose area is proportional to its
  mass.  The open circles in the upper right represent $10^3$ and
  $10^4$ \Ms~star clusters.  The upper histogram shows the SFR.  The
  right histogram depicts the stellar metallicity distribution.  The
  intense halo shows a large spread in metallicity at $z>10$ because
  these stars formed in progenitor halos that were enriched by
  different SN explosions.  At $z<10$, the majority of stellar
  metallicities increase as the halo is self-enriched.  The spikes in
  metallicity at $t$ = 620, 650, and 700 Myr show induced star
  formation with enhanced metallicities in SN remnant shells.  The
  dashed lines in the left panel guide the eye to two stellar
  populations that were formed in two satellite halos, merging at
  $z=7.5$.  The quiet halo evolves in relative isolation and steadily
  increases its metallicity to [Z/H] $\sim$ --2 until there is an
  equilibrium between \textit{in-situ} star formation and metal-poor
  inflows from filaments.}
\end{figure*}


To further demonstrate the metallicity floor in high-redshift dwarf
galaxies, we show the average metallicity from Pop III SNe \zthree~in
all halos as a function of halo mass at $z =$ 7, 8, 9, and 10 in
Fig. \ref{fig:z3}, and it is strongly dependent on halo mass.  The
\zthree~error bars mark the 25\% and 75\% percentiles of the
distribution at $M = M_{\rm halo}$ because the underlying distribution
is not necessarily Gaussian at all halo masses.  Above $3 \times 10^7
\Ms$, all halos enriched to $\zthree \ga -4$ with little time
evolution, further illustrating the metallicity floor.  Below this
halo mass, the distribution is roughly bimodal composed of pristine
and metal-enriched halos.

\subsection{Star formation history}
\label{sec:pop}

The most massive progenitor of the quiet halo interestingly never
hosts a Pop III star.  Instead a nearby halo forms a Pop III star,
which (randomly) produces a PISNe at $z=16$.  The blastwave overruns
the most massive progenitor, and the dense core survives this event
and is enriched by this PISN, triggering the transition to Pop II SF.
Other progenitors host three Pop III stars, forming at $z = 15.4,
14.2, 13.8$, with the latter producing a PISN.  Metal enrichment from
these two PISNe and Pop II SF quench Pop III SF in this halo.  The
progenitors of the intense halo host a total of 56 Pop III stars with
21 producing PISNe and the other 35 producing intermediate-mass BHs.
The first forms in a $6 \times 10^5 \Ms$ halo at $z=19$.  Pop III
stars form on a regular interval in the halo's progenitors until $z=9$
when most of these halos enter the metal-rich bubble surrounding the
intense halo.


Figure \ref{fig:pops} shows the SF history (SFH), metallicity
distribution, and SF rates of both halos.  A nearby PISNe provides a
metallicity floor of [Z/H] = --2.8 in the quiet halo at which
metallicity the first Pop II stars form.  The stellar metallicity
evolution exhibits what is expected from an isolated system with the
stellar feedback steadily enriching the ISM, resulting in a
correlation between stellar age and metallicity.  After $z=10$ the
metallicities plateau at [Z/H] = --2.1 for reasons previously
discussed.  The SFR peaks at $z=10$ and decreases as the cold gas
reservoir is depleted.  Around $z=7.5$, a 25:1 minor merger occurs,
and the gas inside the satellite halo is compressed, triggering
metal-poor, [Z/H] = --3.2, SF during its nearest approach.  This halo
remains metal-poor because most of metal enrichment in the quiet halo
occurs in bi-polar flows perpendicular to the galaxy disk and
filament.  Stars with [Z/H] $<$ --3 compose 1.6 percent of the total
stellar mass.
       
In contrast with the quiet halo, the intense halo undergoes a few
mergers of halos with an established stellar population.  This creates
a superposition of age-metallicity tracks in the SFH, seen in the
complexity of Figure \ref{fig:pops}.  The first two Pop II stellar
clusters have an unexpectedly high metallicity [Z/H] $\sim$ --1, which
occurs when a PISN blastwave triggers SF in two neighboring halos.
Most of the early SF have [Z/H] = --2.5.  At $z=9$, the halo's virial
temperature reaches \tento{4} K.  This combined with a 10:1 merger
creates a starburst that quickly enriches the halo to [Z/H] = --1.5 by
$z=8$.  The halo continues to enrich itself afterwards.  The spikes in
the scatter plot correspond to SN triggered SF in nearby molecular
clouds that are enriched up to a factor of 10 with respect to the ISM.
However their mass fraction are small compared to the total stellar
mass.  The starburst at $z=9$ creates a bimodal metallicity
distribution with peaks at [Z/H] = --2.4 and --1.2 with the metal-rich
component mainly being created after the starburst.  Two systems with
sizable stellar components merge into the halo at $z \sim 8$, and
their stellar populations, which are traced by dashed lines in Figure
\ref{fig:pops}, are still discernible in the metallicity-age plot.
Stars with [Z/H] $<$ --3 compose 1.8 percent of the total stellar
mass.

\subsection{Mass-to-light ratios}


\begin{figure}
\epsscale{1.15}
\plotone{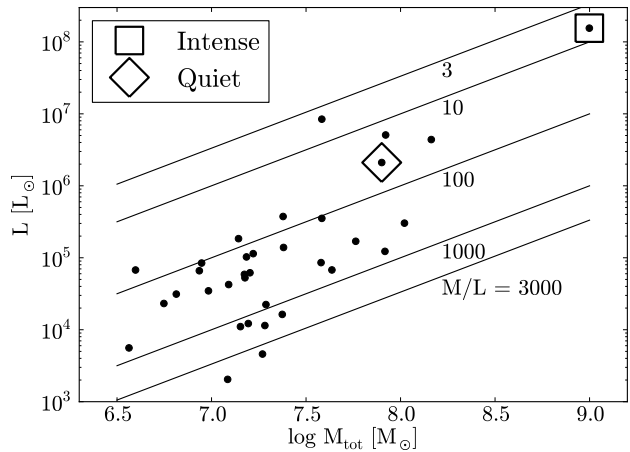}
\epsscale{1}
\caption{\label{fig:massfn} Total Pop II luminosities of halos versus
  total halo mass with constant mass-to-light ratios overplotted at
  redshift 7.  The halos with $M \gsim 10^8 \Ms$ form stars
  efficiently because they can cool through atomic line cooling,
  whereas the lower mass galaxies rely on molecular hydrogen cooling
  only.  The low M/L ratios and spread across two orders of magnitude
  are similar to ones observed in nearby ultra faint dwarf galaxies.}
\end{figure}


Mass-to-light ratios are crucial in interpreting observations when
connecting the galaxy to its DM halo, and ultimately to models of
galaxy formation.  For this paper, we use a simple model to calculate
the total stellar luminosity of each galaxy.  In a later paper, we
will follow up with more sophisticated models to determine more
accurate observables.  Stars in each particle are treated as an
instantaneous burst.  Using the empirical law of main sequence
lifetime $t_{\rm MS} \propto M^{-5/2}$, the maximum stellar mass in
such a system is dependent only on its age $t_\star$, so $M_{\rm up} =
(t_\star / 10\;\mathrm{Gyr})^{-2/5}$.  Then combining another scaling
law $L \propto M^{7/2}$ and the Salpeter IMF $\psi(M) \propto
M^{-\alpha}$,
\begin{eqnarray}
  \label{eqn:lum}
  L/M_\odot &=& \int_{M_{\rm lo}}^{M_{\rm up}} \psi(M) L(M) dM
  \nonumber\\ &\simeq& M_{\rm up}^{9/2-\alpha} (9/2 - \alpha)^{-1},
\end{eqnarray}
where we set $M_{\rm lo} \equiv 0$ because the total luminosity of
low-mass stars is negligible.  Then the galaxy luminosity is the sum
of all star particle luminosities, which we plot in
Fig. \ref{fig:massfn} with respect to halo mass, along with the DM
halo mass function.

Recall that the most massive halo ($M_{\rm tot} = 10^9 \Ms$) is
undergoing a starburst at $z=7$, translating into a relatively high
$M/L = 6$.  The lower mass galaxies show a spread in M/L between 30
and 3000, similar to the Sloan Digital Sky Survey (SDSS) observations
of local ultra faint dwarf (uFd) galaxies \citep[e.g.][]{Strigari08}.
It has been suggested that these low-mass ($M_{\rm tot} \lsim 10^8
\Ms$) galaxies that cannot cool through hydrogen line cooling are the
progenitors of uFd galaxies \citep{Bovill11a, Bovill11b}, where all of
the star formation occurs before reionization and was subsequently
terminated by UV heating from external sources.  These galaxies have
stellar masses ranging from 210 \Ms~(SDSS J1058+2843) to $3.2 \times
10^5$ \Ms~(Dra) with most between $10^3$ and $10^4$
\Ms. \citep{Martin08_uFd}.  Because the total stellar population would
be sampled, on average, by several star particles, we cannot make
robust conclusions about the nature of uFd galaxies.

\section{Discussion and Conclusions}
\label{sec:discuss}

We have focused on the birth of two galaxies prior to reionization
with a cosmological AMR radiation hydrodynamics simulation with very
high spatial (1 pc comoving) and mass resolution (1800 \Ms).  With
this resolution, we are able to follow the Pop III star formation and
radiative and chemical feedback effects in hundreds of minihalos that
contribute to the formation of a protogalaxy.  Supernovae from Pop III
stars provide the necessary heavy elements for the transition to a
Population II stellar population, which we have directly simulated.
The galaxies simulated here represent a population of uncharacterized
high-redshift dwarf galaxies, which should be detected with the
upcoming \textit{James Webb Space Telescope}, that should contribute a
significant fraction of the ionizing photons during reionization.
Recently, a $z=8.6$ spectroscopically confirmed galaxy was found to
reside in an \ion{H}{2} region that extends beyond 1 Mpc.  However,
using the SFRs and \lya~continuum as estimators, this galaxy produces
enough ionizing radiation to create an \ion{H}{2} region between $\sim
0.1(f_{\rm esc}/0.1)^{1/3}$ Mpc and $0.5(f_{\rm esc}/0.1)^{1/3}$ Mpc,
where $f_{\rm esc}$ is the fraction of ionizing radiation escaping
into the IGM \citep{Lehnert10_z8.6}.  This discrepancy could be caused
by undetected satellite dwarf galaxies, such ones presented here, that
contribute to local radiation field, or an ionizing radiation escape
fraction near unity \citep{Wise09, Razoumov10, Paardekooper11,
  Yajima11}.

We find that one PISN is sufficient to enrich the star-forming halo
and surrounding $\sim 5$ kpc to a metallicity of 10$^{-3} Z_\odot$,
given $M_{\rm char} = 100~\Ms$.  DLA systems have a metallicity floor
on the same order \citep{Wolfe05_Review, Penprase10}, and metal
enrichment from Pop III SNe provides a possible explanation
\citep{Kobayashi11}.  Our simulation strengthens this claim of a
metallicity floor from Pop III SNe \citep[see also][]{Tornatore07,
  Karlsson08, Maio11_Enrich}, where our work improves on previous
studies by resolving star-forming molecular clouds and employing more
realistic feedback physics, in particular, radiation transport.

If the first stars have a lower characteristic mass that favor
hypernovae \citep{Tumlinson07_IMF}, then this metallicity floor should
be lowered by a factor of $\sim 10$ because the metal ejecta from a 40
\Ms~hypernova is 8.6 \Ms~\citep{Nomoto06}, compared to 85 \Ms~of
metals produced by a 180 \Ms~PISN \citep{Heger02}.  Both explosions
have approximately $3 \times 10^{52}$ erg.  Therefore the blastwaves
from these two explosions should expand to equal radii before stalling
in the IGM.  We note that stronger metal-line cooling in the PISN case
may cause the blastwave to stall at a marginally smaller radius.  

Building upon this idea of a lower metallicity floor, consider a
pre-galactic halo that has hosted a Pop III star, producing a SN.  The
majority of the gas is ejected into the IGM and re-accretes after
$\sim 10-50$ Myr \citep{Johnson07, Wise08_Gal}.  What happens if the
metallicity is not greater than the critical metallicity?  Massive Pop
III stars would form again in this halo; however, enriching the
surrounding IGM with Pop III SNe and ejecting material becomes more
difficult in halos with $M \gsim 10^7 \Ms$ \citep{Whalen08_SN}.  Thus
it is more likely for these halos to retain more of its gas during
star formation episodes and self-enrich its host halo and less of the
IGM.  This would accelerate the transition from Pop III to II in more
massive halos.

We conclude that it only takes one, at most two, SNe from Pop III
stars in the halo progenitors to complete the transition to Population
II \citep{Frebel10}.  However, this does not prohibit more than two
instances of Pop III star formation in halo from happening because the
stars can produce BHs with little or no metal ejecta, keeping the
surrounding environment pristine.  Even though we find that Pop III
SNe create a floor of $\sim 10^{-3} \zsun$, gas with lower
metallicities can still exists in nearby halos, where the metal-rich
SN ejecta will mix slowly into dense cores as the blastwave overtakes
them \citep{Cen08}.  These environments may be the host of true
extremely metal-poor star formation, such as the recently discovered
halo star with all detected metals below $10^{-4.2} \zsun$
\citep{Caffau11}.

We also find that the merger history plays an important role in
supplying gas into halos after its first epoch of SF.  Mergers of
halos below the filtering mass are inefficient in providing gas
whereas the opposite is true for merging halos above the filtering
mass.  Halos do not necessarily need $\tvir \ge 10^4$ K to form
significant stellar populations; however SFRs dramatically increase,
and thus metal enrichment, when this threshold is reached.  Halos with
mostly gas-poor mergers or a quiescent merger history result in a
monotonic increase in metallicity with time.  SFHs become more complex
with multiple metallicity-age tracks if the halo experiences mergers
with other halos that have an established stellar population.
Furthermore SF that is triggered by blastwaves interacting with
molecular clouds can have metallicities up to a factor of 10 higher
than the main starburst.  Starbursts at $\tvir = 10^4$ K can enrich
the host halo enough to create a bimodal metallicity distribution,
where the metal-poor component is created before the burst.  Note that
mergers of stellar populations can also create a similar bimodal
distribution, which are observed in dSphs \citep{Battaglia11}.

Because our simulation only samples 1 comoving Mpc$^3$, it misses the
more rare density fluctuations and is biased toward late galaxy
formation \citep{Barkana04}.  In such rare galaxies, the physical
scenario described by this work is largely unchanged but is shifted to
higher redshifts.  Thus our results can be generalized to any epoch
prior to cosmic reionization.  However, two possible consequences in
very high ($z \gsim 20$) redshift galaxy formation are 1) an
intermediate mass IMF \citep{Larson98, Tumlinson07_IMF,
  2009ApJ...691..441S}, resulting from the CMB temperature floor,
could be more prevalent, and 2) streaming gas velocities arising from
the recombination epoch \citep{Tselia10} could suppress star formation
in halos below $10^6 \Ms$ \citep{Tselia10_Minihalo, Greif11_Delay}.

We have shown that it is possible to simulate the formation of a
high-redshift dwarf galaxy and its entire SFH with radiative and
mechanical feedback.  These results provide invaluable insight on the
first galaxies and the role of metal-free stars in the early universe.
We plan to follow up this work by focusing on observational
connections with high-redshift galaxies and Local Group dwarf galaxies
in the near future that can better constrain the physical models and
assumptions, such as the critical metallicity, radiation backgrounds,
and IMFs, used in \textit{ab initio} dwarf galaxy simulations.

\acknowledgments

JHW thanks Renyue Cen, Amina Helmi, Marco Spaans, and Eline Tolstoy
for enlightening discussions.  We thank an anonymous referee for
helpful suggestions.  JHW is partially supported by NASA through
Hubble Fellowship grant \#120-6370 awarded by the Space Telescope
Science Institute, which is operated by the Association of
Universities for Research in Astronomy, Inc., for NASA, under contract
NAS 5-26555.  MJT and MLN acknowledge partial support from NASA ATFP
grant NNX08AH26G.  MJT~acknowledges support by the NSF CI TraCS
fellowship award OCI-1048505.  Computational resources were provided
by NASA/NCCS award SMD-09-1439.  TA acknowledges financial support
from the {\em Baden-W\"{u}rttemberg-Stiftung} under grant
P-LS-SPII/18, the Heidelberg Institut f\"ur Theoretische Studien. This
work was partially supported by NASA ATFP grant NNX08AH26G, NSF
AST-0807312, and NSF AST-1109243. This research has made use of NASA’s
Astrophysics Data System Bibliographic Services.  The majority of the
analysis and plots were done with \texttt{yt} \citep{yt_full_paper}.


\end{document}